\documentclass[aip,graphicx,
 amsmath,amssymb,
 reprint,%
Conference Proceedings]{revtex4-1}

\usepackage[utf8]{inputenc}
\usepackage{graphicx}
\graphicspath{ {./fig/} }
\usepackage{bm}
\usepackage{hyperref}

\usepackage[dvipsnames]{xcolor}
\DeclareGraphicsExtensions{.pdf,.jpg}

\usepackage[normalem]{ulem}

\DeclareUnicodeCharacter{2212}{-}

\draft 

\begin{document}


\title{Effects of Rim Fluctuations in Classical Nucleation Theory of Virus Capsids} 



\author{Alexander Bryan Clark}
\affiliation{Department of Physics \& Astronomy, University of California, Riverside CA 92521, USA}

\author{Paul van der Schoot}
\affiliation{Department of Applied Physics and Science Education, Eindhoven University of Technology, Postbus 513, 5600 MB Eindhoven, Netherlands}

\author{Henri Orland}
\affiliation{Institut de Physique Th\'eorique, CNRS, CEA, Universit\'e Paris-Saclay, Gif-Sur-Yvette, France}
\affiliation{Department of Physics, School of Sciences, Great Bay University, Dongguan, Guangdong 523000, China}

\author{Roya Zandi}
\email{roya.zandi@ucr.edu}
\affiliation{Department of Physics \& Astronomy, University of California, Riverside CA 92521, USA}

\date{\today}

\begin{abstract}
Most spherical viruses exhibit icosahedral symmetry, yet the growth of viral shells remains poorly understood due to the short lifetimes and broad size distribution of assembly intermediates. Classical nucleation theory has been widely applied to describe this process, but it treats the boundary of a growing shell as rigid and structureless. Here, we extend classical nucleation theory by incorporating thermal fluctuations of the capsid rim using both discrete and continuum descriptions. Allowing the rim of a partially formed capsid to undergo small geometric undulations, we show that these fluctuations generate an entropic contribution that renormalizes the effective line tension. As a result, rim fluctuations can either promote or hinder capsid closure, depending on the subunit–subunit binding free energy, temperature, and fluctuation amplitude. We find that fluctuations generally lower the nucleation barrier when the binding free energy is below a threshold value, while for sufficiently strong binding they can instead raise the barrier by stabilizing incomplete capsids through a finite-size entropy penalty associated with rim closure. By moving beyond the idealized capillarity approximation, our results provide a controlled extension of classical nucleation theory that clarifies how boundary fluctuations influence capsid nucleation and growth.
\end{abstract}



\maketitle

\section{Introduction}
Viral capsids provide a canonical example of molecular self-assembly, forming robust, highly symmetric shells from many identical protein subunits. Their geometry and reproducibility have long made them model systems for studying fundamental principles of self-organization. Early work by Crick and Watson and by Caspar and Klug established the concept of quasi-equivalence and the Caspar–Klug classification, explaining how identical subunits assemble into stable spherical shells, most commonly with icosahedral symmetry \cite{crick1956structure, klug1961structure, Rossmann_2013,Zandi2020,Hagan2010,Siber2008,Gonca2014,Gonca2016,Li2018,Hagan2014_ch}. Recent coarse-grained simulations have incorporated quasi-equivalence by allowing interaction-triggered conformational switching, which enables the spontaneous assembly of icosahedral capsids around genomes\cite{,Li2025}. Owing to their high symmetry and relative simplicity, icosahedral capsids remain paradigmatic systems for theoretical studies of assembly \cite{Twarock_Icosahedral_design, Robijn_2005_Elasticity_Theory, bancroft_self_assembly_1970,Elrad2008}.

Classical nucleation theory (CNT) has been widely used to describe capsid assembly as a thermally activated process governed by a competition between bulk free-energy gain and an interfacial free energy penalty associated with the open rim of a partial shell \cite{zandi_classical_2006, ZLOTNICK2003269, Li2018,panahandeh2020virus}. Within this framework, assembly proceeds by overcoming a free-energy barrier associated with a critical nucleus, whose size and height depend on the temperature, protein concentration, and subunit binding free energy. 
CNT successfully captures many qualitative features of capsid assembly kinetics and thermodynamics. However, simulations and experiments indicate that assembly pathways often involve disordered or defect-containing intermediates that subsequently reorganize into closed shells, implying that the boundary of a growing capsid is dynamically remodeled during growth \cite{Hagan2010,Li2025}.  

A key idealization of CNT is the assumption that the rim of a partially assembled capsid is a rigid, circular boundary with a fixed line tension \cite{Sear_2007, zandi_classical_2006}. In reality, the rim is composed of discrete protein subunits and may undergo thermal fluctuations that modify its energetic and entropic contributions. Previous analytical and computational studies suggest that interfacial fluctuations and elastic properties can significantly influence assembly pathways and nucleation barriers \cite{lidmar2003virus, PhysRevE.81.041925, dePablo2013, KOL20071777}. Even small deviations from a circular rim can raise or lower the barrier height, depending on how rim geometry couples to subunit interactions \cite{Stefan}, motivating a more explicit treatment of rim flexibility.

In this work, we extend classical nucleation theory by explicitly incorporating thermal fluctuations of the capsid rim into the free-energy landscape of partially formed shells. We allow the rim to undergo small geometric undulations and show that these fluctuations provide an entropic contribution that renormalizes the effective line tension. As a result, rim fluctuations generally lower the nucleation barrier and shift the critical nucleus size relative to the rigid-rim CNT prediction. Importantly, we find a dual role of fluctuations: while they typically promote assembly by reducing the barrier, in regimes of very strong subunit binding or very small intermediates, the entropy associated with enforcing rim closure can produce a modest increase in the barrier. This non-monotonic behavior arises from the combined effect of subunit interaction energy and a finite-size entropy correction imposed by the closure constraint.

To demonstrate the robustness of these conclusions, we develop two complementary descriptions of rim fluctuations: (i) a discrete step model in which each subunit contributes upward or downward rim deformations subject to an exact closure condition, and (ii) a continuum capillary-wave model treating the rim as a fluctuating elastic ring. Both approaches consistently show that rim fluctuations modify the free-energy profile by reducing the effective line tension, whereas the closure constraint introduces a non-extensive contribution that increases the barrier. 

Overall, our results provide a controlled extension of classical nucleation theory that explicitly couples boundary geometry to assembly thermodynamics. By moving beyond the rigid-rim approximation, we clarify how entropic contributions from rim fluctuations reshape the nucleation free-energy landscape and influence the earliest stages of capsid assembly.

\section{Discrete Rim Fluctuations}
In this section, we introduce a discrete microscopic model to describe thermal fluctuations of the rim of a partially assembled capsid and their effect on the free energy of assembly. The model captures how small, local deformations of the rim modify the nucleation landscape and provides a reference for the continuum description developed later.

\begin{figure}
    \centering
    \includegraphics[width=0.99\linewidth]{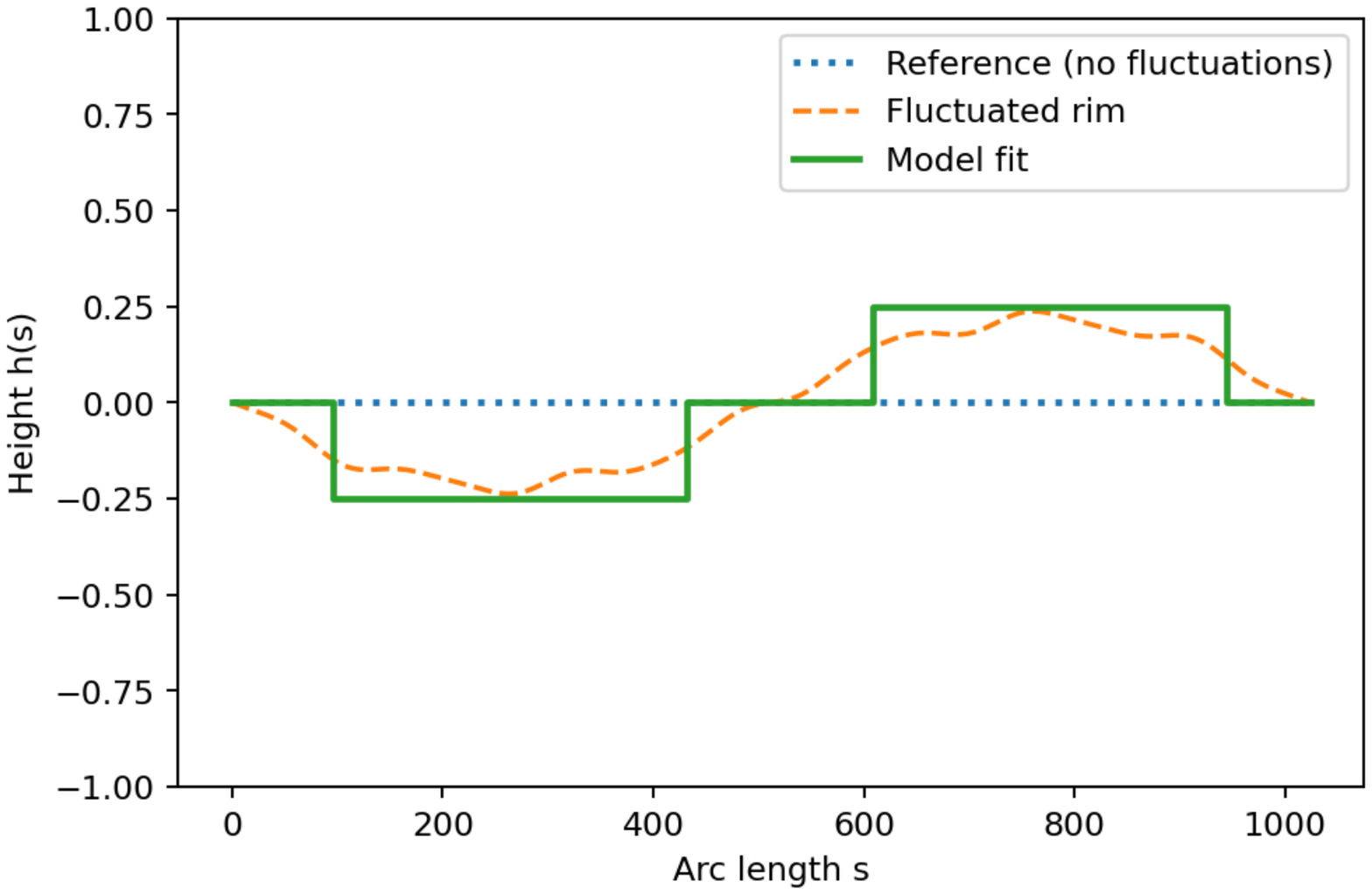}
    \caption{A comparison of an actual rim fluctuation to the discrete model. A randomly generated rim height profile $h(s/L)$, with maximum deviation $|b|=0.25$, along a rim segment of arc length $s$ (orange dashed line) is compared to the best-fit three-state discrete step model (solid green line). The blue dotted line indicates a reference rim with no fluctuations.}
    \label{discretized_arc_length_fit}
\end{figure}

The rim is modeled as a circular boundary composed of discrete subunits and characterized by an effective number of rim bonds $N$, which provides a geometric measure of the rim length. Each subunit may undergo a small vertical displacement relative to a reference circular contour, representing local rim fluctuations within the harmonic regime. A comparison between a continuous rim deformation and its representation in the discrete step model is shown in Figure~\ref{discretized_arc_length_fit}.

The energy of a fluctuating rim is
\begin{align}
    E = \gamma \sum_{i=1}^{N} \sqrt{a^2 + b^2 n_i^2},
\end{align}
where $\gamma$ is the line tension, $a$ is the horizontal length per subunit, $b$ is the vertical step height, and $n_i \in \{0,\pm1\}$ denotes an upward, downward, or zero step at bond $i$. Closure of the rim requires the sum of step heights to be zero, $\sum_{i=1}^{N} n_i = 0$. The corresponding configurational partition function is
\begin{align}
    Z = \sum_{\{n_i\}} \exp\!\left[-\beta \gamma \sum_{i=1}^{N} \sqrt{a^2 + b^2 n_i^2}\right]\delta_{\sum_{i=1}^{N} n_i,0}\label{discrete Z},
\end{align}
with $\beta = (k_{\textsc{b}}T)^{-1}$. Using the integral representation of the Kronecker delta function, $\delta\!\left(\sum_i n_i\right)= \frac{1}{2\pi}\int_{-\pi}^{\pi} \,\mathrm{d}k \exp\!\left[i k \sum_i n_i\right]$, and factoring out the bulk contribution, the partition function can be written as
\begin{align}
    Z &= e^{-\beta \gamma N a}\int_{-\pi}^\pi \frac{\mathrm{d}k}{2\pi}\left(1 + 2\cos(k)\, e^{-\beta \gamma (c-a)}\right)^N,
\label{factored Z}
\end{align}
where $c=\sqrt{a^2+b^2}$. For large $N$, the dominant contribution to the integral comes from the small-$k$ region, $k \sim 0$. 
This saddle-point evaluation is strictly accurate only if 
\begin{align}
    N\frac{2e^{-\beta \gamma (c-a)}}{1+2e^{-\beta\gamma (c-a)}} \gg 1,
\end{align}
which translates to $T\gtrsim \gamma (c-a)/k_{\textsc{b}}\ln N$; otherwise the strict low-temperature limit should be taken from Eq.~\eqref{discrete Z} (see Appendix~\ref{appendix D}). Expanding $\cos k \approx 1 - k^2/2$, defining 
\begin{equation}
    \varepsilon^{\text{D}} =  \gamma \Delta \ell>0 \label{DeltaE}
\end{equation}
where the change in rim length is given as $\Delta \ell=c-a$, and evaluating the resulting Gaussian integral yields the free energy via $\beta F = -\ln Z$,
\begin{align} \label{discrete F}
    \beta F = \beta \gamma N a - N \ln\!\left[1 + 2e^{-\beta \varepsilon^{\text{D}}}\right] + \frac{1}{2}\ln\!\left[\frac{2\pi N}{1 + \tfrac{1}{2} e^{\beta \varepsilon^{\text{D}}}}\right].
\end{align}
The first term describes the contribution from a rigid circular rim. The second term is an extensive correction arising from rim fluctuations, while the final logarithmic term is a non-extensive correction associated with the closure constraint.

\subsection{Properties of Three-State Rim Fluctuations}
The dimensionless free energy in Eq.~\eqref{discrete F} allows us to determine whether rim fluctuations lower or raise the nucleation barrier. To identify when the fluctuation contribution changes sign, we define the fluctuation-induced free-energy correction as
\begin{align}
    \Delta F = F - F_0 ,
    \label{eq:deltaF_def}
\end{align}
where $F_0=\gamma Na$ denotes the free energy of a rigid rim without fluctuations. Setting $\Delta F = 0$ yields the condition
\begin{align}
    2\pi N
    =
    \left(1+\tfrac{1}{2}e^{\beta \varepsilon^{\text{D}}}\right)
    \left(1+2e^{-\beta \varepsilon^{\text{D}}}\right)^{2N},
    \label{eq:dis_zero_cond}
\end{align}
which determines the boundaries separating regimes where rim fluctuations increase or decrease the free energy.  It is convenient to introduce
\begin{align}
    A &= \ln\!\left(1+2e^{-\beta \varepsilon^{\text{D}}}\right), \nonumber\\
    B &= 1+\tfrac{1}{2}e^{\beta \varepsilon^{\text{D}}},
    \label{eq:AB_def}
\end{align}
which are both positive. Equation~\eqref{eq:dis_zero_cond} can then be written compactly as
\begin{align}
    2\pi N = B\, e^{2 N A},
    \label{eq:prelambert}
\end{align}
which can be solved analytically using the Lambert $W$ function, defined implicitly by
\begin{align}
    y e^{y} = x
    \quad \Longleftrightarrow \quad
    y = W(x).
    \label{eq:lambert_definition}
\end{align}
The Lambert $W$ function has two real branches for arguments in the interval $[-1/e,0)$: the principal branch $W_0$, which takes values in $[-1,0)$, and the lower branch $W_{-1}$, which takes values in $(-\infty,-1]$. At the boundary value $x=-1/e$, both branches coalesce at $W=-1$; for $x<-1/e$, no real solutions exist.

Applying Eq.~\eqref{eq:lambert_definition} to Eq.~\eqref{eq:prelambert}, and labeling the two solutions by the branch index $k\in\{0,-1\}$, yields
\begin{align}
    N_k
    =
    -\frac{1}{2A}
    W_k\!\left(-\frac{A B}{\pi}\right),
    \label{eq:discrete_lambert_solution}
\end{align}
with $N_0 < N_{-1}$. These two values of $N$ correspond precisely to the rim sizes at which $\Delta F = 0$, i.e., the boundaries separating regimes where rim fluctuations increase or decrease the free energy.

\begin{figure}
    \centering
    \includegraphics[width=0.99\linewidth]{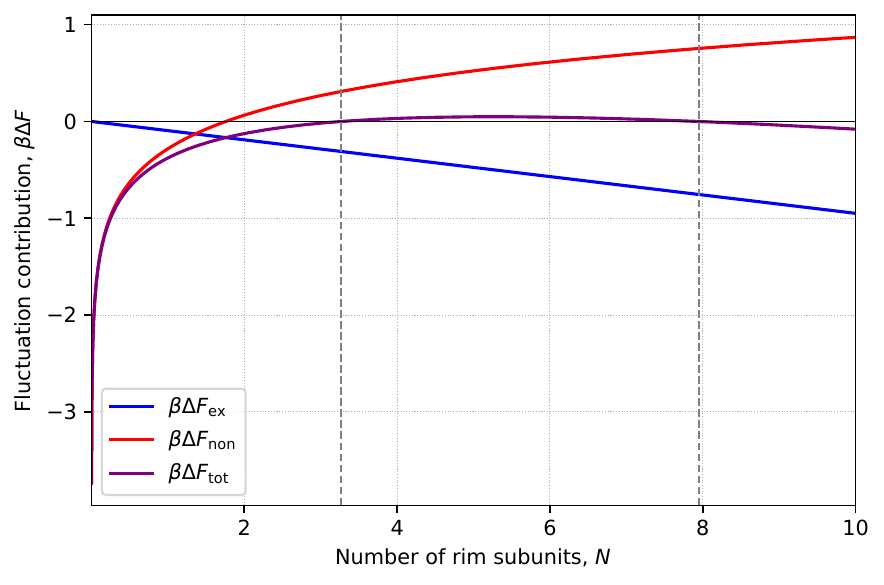}%
    \caption{Contributions to the dimensionless fluctuation free energy $\beta \Delta F$ from Eq.~\eqref{eq:deltaF_def} are shown as a function of the number of rim subunits $N$ for $\beta \varepsilon^{\mathrm{D}} = 3 > \beta \varepsilon^{\mathrm{D}}_\ast$. The blue curve represents the extensive contribution $\beta \Delta F_{\mathrm{ext}}$, while the red curve represents the non-extensive contribution $\beta \Delta F_{\mathrm{non}}$ arising from the closure constraint. The purple curve shows the total fluctuation correction $\beta \Delta F$. The thin vertical dashed gray lines mark $N_0$ (principal Lambert branch, left) and $N_{-1}$ (lower Lambert branch, right), the two values of $N$ for which $\Delta F = 0$ according to Eq.~\eqref{eq:discrete_lambert_solution}. As a result, $\beta \Delta F > 0$ only within the finite interval $N_0 < N < N_{-1}$.}
    \label{discrete Delta F less than plot}
\end{figure}

Real solutions exist only when $AB \le \pi/e$, which defines a critical value 
$\beta \varepsilon^{\mathrm{D}}_\ast \approx 1.756$. For $\beta \varepsilon^{\mathrm{D}} < \beta \varepsilon^{\mathrm{D}}_\ast$, the fluctuation contribution $\beta \Delta F$ is strictly negative for all $N$, implying that rim fluctuations always lower the free energy. In contrast, for $\beta \varepsilon^{\mathrm{D}} > \beta \varepsilon^{\mathrm{D}}_\ast$, Eq.~\eqref{eq:discrete_lambert_solution} admits two real solutions, defining a finite window of intermediate rim sizes for which $\beta \Delta F > 0$. This behavior is illustrated in Fig.~\ref{discrete Delta F less than plot}, where the fluctuation correction becomes positive only over a bounded range of $N$.

The maximum in Fig.~\ref{discrete Delta F less than plot} arises from the competition between two contributions with distinct scaling in $N$. The extensive component of the fluctuation free energy grows linearly with $N$ and always lowers the free energy, while the non-extensive contribution associated with the closure constraint grows sublinearly and raises it. As a result, $\Delta F(N)$ is a concave function of $N$. Setting $\mathrm{d}(\Delta F)/\mathrm{d}N = 0$ yields a maximum at $N_{\mathrm{max}} \simeq 1/(2A)$, where the closure-induced penalty balances the extensive contribution. 
While the extensive entropy from rim fluctuations lowers the free energy and favors assembly, the closure constraint suppresses fluctuations in small clusters and introduces a finite-size penalty. When subunit--subunit binding is sufficiently strong or the critical nucleus is small, this finite-size penalty can temporarily outweigh the extensive gain, producing a slight increase in the nucleation barrier.

In the limit of no geometric rim deformation, $ \Delta \ell = 0$, one finds $A B = (3/2)\ln 3 > \pi/e$, and consequently $\beta \Delta F$ remains negative for all $N$. In this case, geometric rim fluctuations vanish, but the discrete model still exhibits a finite configurational degeneracy associated with the labeling of subunit states. This produces a constant entropic offset relative to the rigid reference free energy and does not correspond to physical rim fluctuations.

To clarify the origin and parameter dependence of this behavior, we now turn to a continuum description of rim fluctuations, where the same Lambert-equation structure reappears and allows for a more detailed analytical classification of fluctuation regimes.

A straightforward generalization of the discrete three-state model to an arbitrary number of states is presented in Appendix~\ref{Appendix A}. Although it does not lead to qualitatively new behavior, it allows a straightforward derivation of the continuous model.

\section{Continuum Limit}
Having examined the three-state discretized model, we now turn to a fully continuum description of the capsid rim. Most classical formulations of capsid nucleation assume that subunits assemble into an approximately spherical geometry, with the interface of a partially assembled structure treated as a continuous circular rim. In this section, we address this scenario by analyzing a capillary-wave model for the rim, in which thermal fluctuations are governed by an effective line tension, in the continuum limit and comparing its predictions with our discrete results.

We begin with a discrete energy functional for a spherical cap whose interface consists of $N$ subunits at the rim. Let $h_j$ denote the height deviation of bond site $j$ relative to the spherical cap, with the reference configuration given by $h_j = 0$ for $j = 1,\ldots,N$. We further assume that fluctuations are measured relative to the existing rim and that site~1 is anchored at $h_1 = 0$. Here, $a$ denotes the characteristic subunit spacing along the rim and $b$ is the corresponding vertical height scale, as described in the previous section. For small slopes, the energy of a configuration $\{h_j\}$ may be approximated by a quadratic functional. Introducing the dimensionless slope parameter $\lambda = b/a$, one can derive a discrete Hamiltonian analogous to that of a harmonic chain. Denoting $n_j = h_j/b$ as the dimensionless height variable at site $j$, the energy can then be expanded to second order in $n_j$ and in the discrete spatial differences between neighboring sites $n_j$ and $n_{j+1}$.
The resulting effective Hamiltonian has the form of a discretized Gaussian surface,
\begin{align}
    H[n] = \gamma a \sum_{j=1}^N \sqrt{1 + \lambda^2 \left( n_{j+1} - n_j \right)^2 } .
\end{align}
For small slopes ($\lambda \ll 1$), 
expanding the square root to quadratic order and neglecting higher-order terms yields
\begin{align}
    H[n] \approx \gamma N a + \frac{\gamma a \lambda^2}{2} \sum_{j=1}^N \left( n_{j+1} - n_j \right)^2 ,
    \label{classic hamiltonian}
\end{align}
where the first term corresponds to the baseline energy of a flat interface of length $Na$, and the second term represents the leading-order energetic cost of rim fluctuations. 

The corresponding partition function is constructed by constraining site~1 to zero height and weighting configurations by the Gaussian energy derived above. Fixing the height of site~1 by imposing the constraint $n_1 = 0$ removes the zero mode associated with rigid vertical translations of the rim and thereby fixes the overall reference configuration. Separating out the baseline energy, the partition function reads
\begin{align}
    Z = e^{-\beta \gamma N a}
    \prod_{j=1}^{N} \int_{-\infty}^\infty \mathrm{d}n_j \,
    \exp\!&\left[
        -\frac{\beta \varepsilon^{\text{C}}}{2}
        \sum_{j=1}^{N} \left( n_{j+1} - n_j \right)^2
    \right]\nonumber\\
    &\times \delta(n_1)\,\delta(n_{N+1}),
    \label{continuum Z}
\end{align}
where the delta functions enforce the circular closure of the rim, and we have defined a continuum elastic energy scale as
\begin{align}
    \varepsilon^{\text{C}}= \gamma a \lambda^2.
\end{align}
Equation~\eqref{continuum Z} may be rewritten by explicitly integrating out the constrained degree of freedom,
\begin{align}
    Z = e^{-\beta \gamma N a}
    \prod_{j=2}^{N} \int_{-\infty}^{+\infty}  \mathrm{d}n_j \,
    \exp\!\left[
        -\beta \varepsilon^{\text{C}}
        \sum_{k,l=2}^{N} n_k M_{kl} n_l
    \right],
\end{align}
where $M_{kl}$ are the elements of an $(N-1)\times(N-1)$ matrix representing the discrete Laplacian on the ring (see Appendix~\ref{appendix continuous} for the explicit form of the matrix). The resulting Gaussian integral can be evaluated analytically, yielding
\begin{align}
     Z &= e^{-\beta \gamma Na}N^{-1/2}\left(\frac{2\pi}{\beta \varepsilon^{\text{C}}}\right)^{(N-1)/2}
\end{align}

Following the procedure outlined in Appendix~\ref{appendix continuous}, the corresponding dimensionless free energy becomes
\begin{align}
    \beta F
    = \beta \gamma N a
    - \frac{N-1}{2}
    \ln\!\left(
        \frac{2\pi}{\beta \varepsilon^{\text{C}}}
    \right)
    + \frac{1}{2} \ln N ,
    \label{continuum approx result}
\end{align}
where the first term, as before, represents the energy of a flat interface of length $N a$, while the remaining terms arise from Gaussian capillary-wave fluctuations of the rim and from the non-extensive correction associated with the closure constraint. 

\subsection{Properties of Continuous Fluctuations}
In the continuum limit, the fluctuation contribution to the free energy, $\Delta F$, is a function of the rim size $N$ and the dimensionless parameter $\beta \varepsilon^{\text{C}}$. Explicitly, the dimensionless fluctuation free energy is given by
\begin{align}
\beta \Delta F
&=
\frac{N-1}{2}
\ln\!\left(
\frac{\beta \varepsilon^{\text{C}}}{2\pi}
\right)
+ \frac{1}{2} \ln N ,
\label{eq:cont_deltaF}
\end{align}
which, as in the discrete model, can be either positive or negative depending on $N$.

For a fixed value of $\beta \varepsilon^{\text{C}} > 0$, the boundaries separating regimes where fluctuations increase or decrease the free energy are obtained by setting $\beta \Delta F = 0$. This yields
\begin{align}
\ln N
&=
(N-1)
\ln\!\left(
\frac{2\pi}{\beta \varepsilon^{\text{C}}}
\right).
\label{eq:cont_zero_condition}
\end{align}
Exponentiating both sides gives
\begin{align}
N
&=
\left(
\frac{2\pi}{\beta \varepsilon^{\text{C}}}
\right)^{N-1},
\label{eq:cont_exponentiated}
\end{align}
and rearranging this expression leads to
\begin{align}
N
\ln\!\left(
\frac{\beta \varepsilon^{\text{C}}}{2\pi}
\right)
\exp\!\left[
N
\ln\!\left(
\frac{\beta \varepsilon^{\text{C}}}{2\pi}
\right)
\right]
&=
\frac{\beta \varepsilon^{\text{C}}}{2\pi}
\ln\!\left(
\frac{\beta \varepsilon^{\text{C}}}{2\pi}
\right),
\label{eq:cont_lambert_form}
\end{align}
which has the same structure as the Lambert equation defined in Eq.~\eqref{eq:lambert_definition}. As described there, real solutions correspond to the two real branches $k \in \{0,-1\}$,
\begin{align}
    N_k=\frac{1}{\ln\left(\frac{\beta \varepsilon^{\text{C}}}{2\pi}\right)}W_k\left(\frac{\beta \varepsilon^{\text{C}}}{2\pi}\ln\left(\frac{\beta \varepsilon^{\text{C}}}{2\pi}\right)\right)
    \label{eq:cont_lambert_solutions}
\end{align}
with $N_0 < N_{-1}$, where $N_0$ and $N_{-1}$ are again the rim sizes at which $\Delta F=0$.

\begin{figure}
    \centering
    \includegraphics[width=0.99\linewidth]{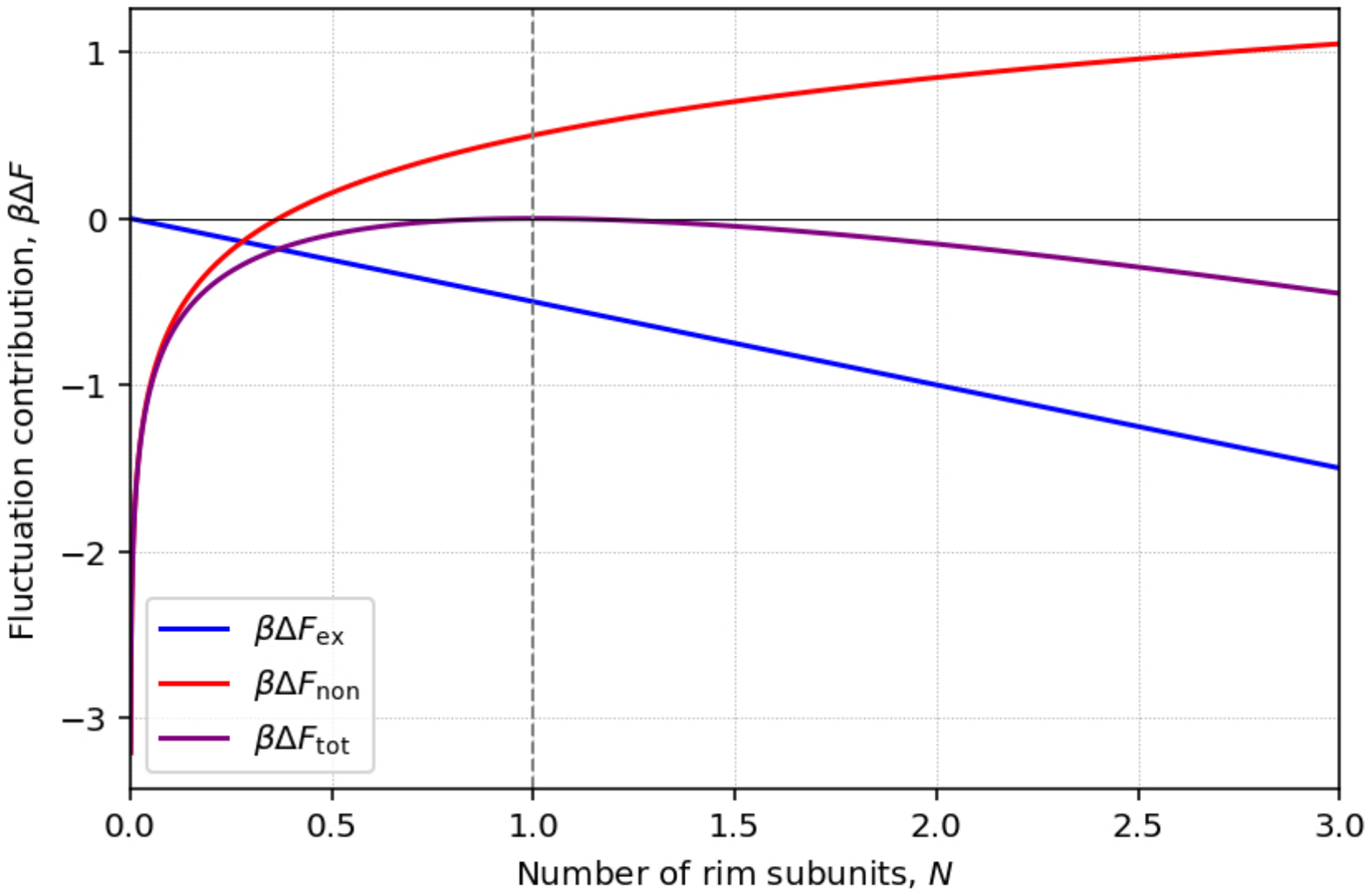}
    \caption{Dimensionless fluctuation free-energy contributions in the continuum rim model from Eq.~\eqref{eq:cont_deltaF} as a function of the rim size $N$ at the threshold parameter value $\beta \varepsilon^{\text{C}}/2\pi=1/e$. The solid blue curve shows the extensive contribution $\beta \Delta F_{\mathrm{ext}}$, the solid red curve shows the non-extensive contribution $\beta \Delta F_{\mathrm{non}}$ arising from the closure constraint, and the solid purple curve shows the total fluctuation correction $\beta \Delta F$. The dashed vertical gray line at $N=1$ marks the unique solution of the zero-condition equation, where the two Lambert branches coalesce and $\beta \Delta F=0$. For all other values of $N$, the total fluctuation correction is negative, indicating that rim fluctuations lower the free energy. Note that only the part $N\geq1$ is physical (or relevant).}
    \label{fig:case_1}
\end{figure}

\begin{figure}
    \centering
    \includegraphics[width=0.99\linewidth]{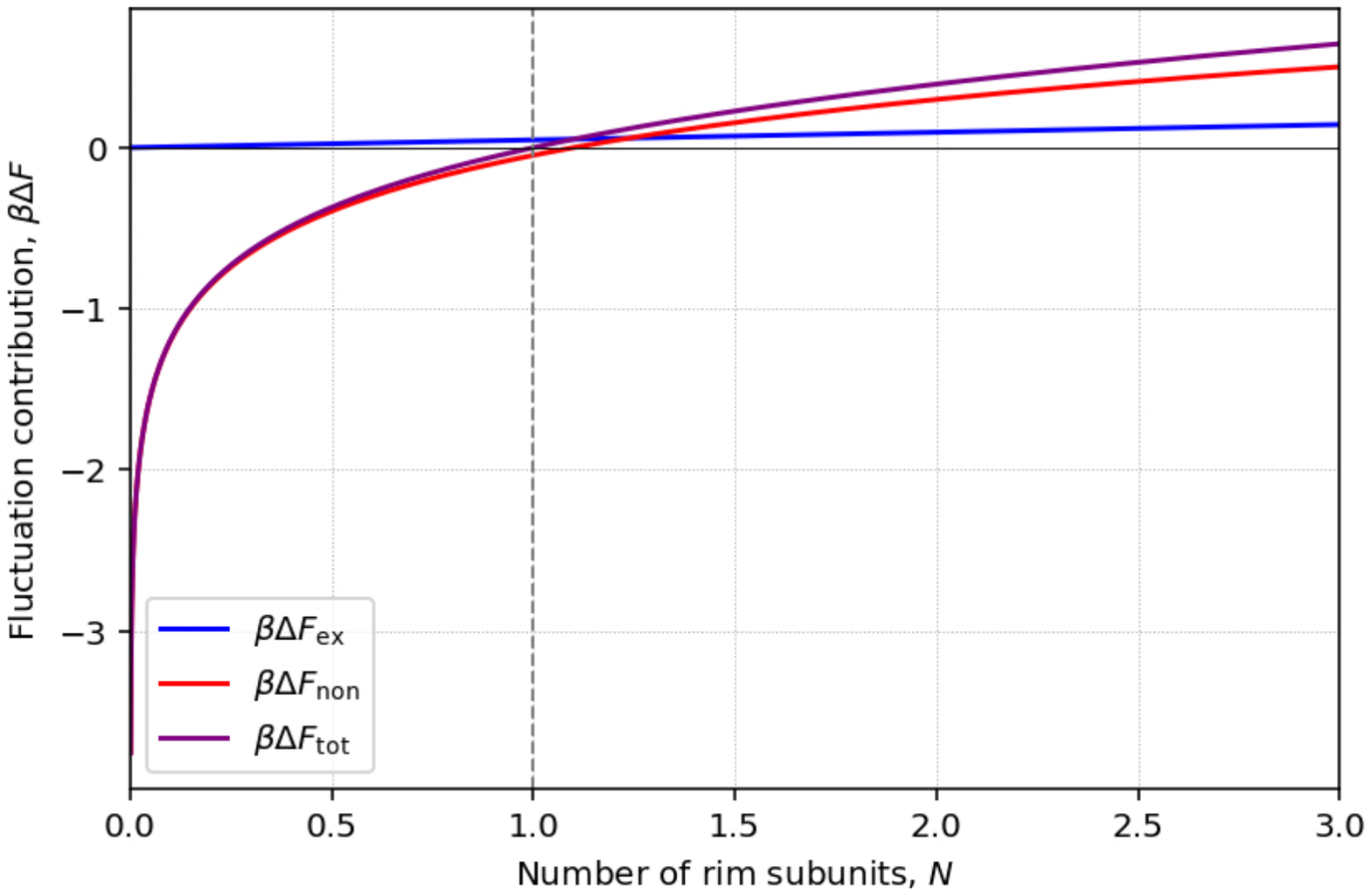}
    \caption{Continuum rim fluctuation contributions to the dimensionless free energy $\beta \Delta F$ from Eq.~\eqref{eq:cont_deltaF} as a function of the rim size $N$ for the parameter value $\beta \varepsilon^{\text{C}}/2\pi=1.1>1/e$. The solid blue curve shows the extensive contribution $\beta \Delta F_{\mathrm{ext}}$, the solid red curve shows the non-extensive contribution $\beta \Delta F_{\mathrm{non}}$, and the solid purple curve shows the total fluctuation correction $\beta \Delta F$. The dashed vertical gray line at $N=1$ corresponds to the single real solution of the zero-condition equation. In this regime, the total fluctuation correction remains positive for all $N>1$, indicating that rim fluctuations uniformly raise the free-energy barrier. The regime $N<1$ is unphysical, as the rim must consist of at least one bond.}
    \label{fig:case_2}
\end{figure}

\begin{figure}
    \centering
    \includegraphics[width=0.99\linewidth]{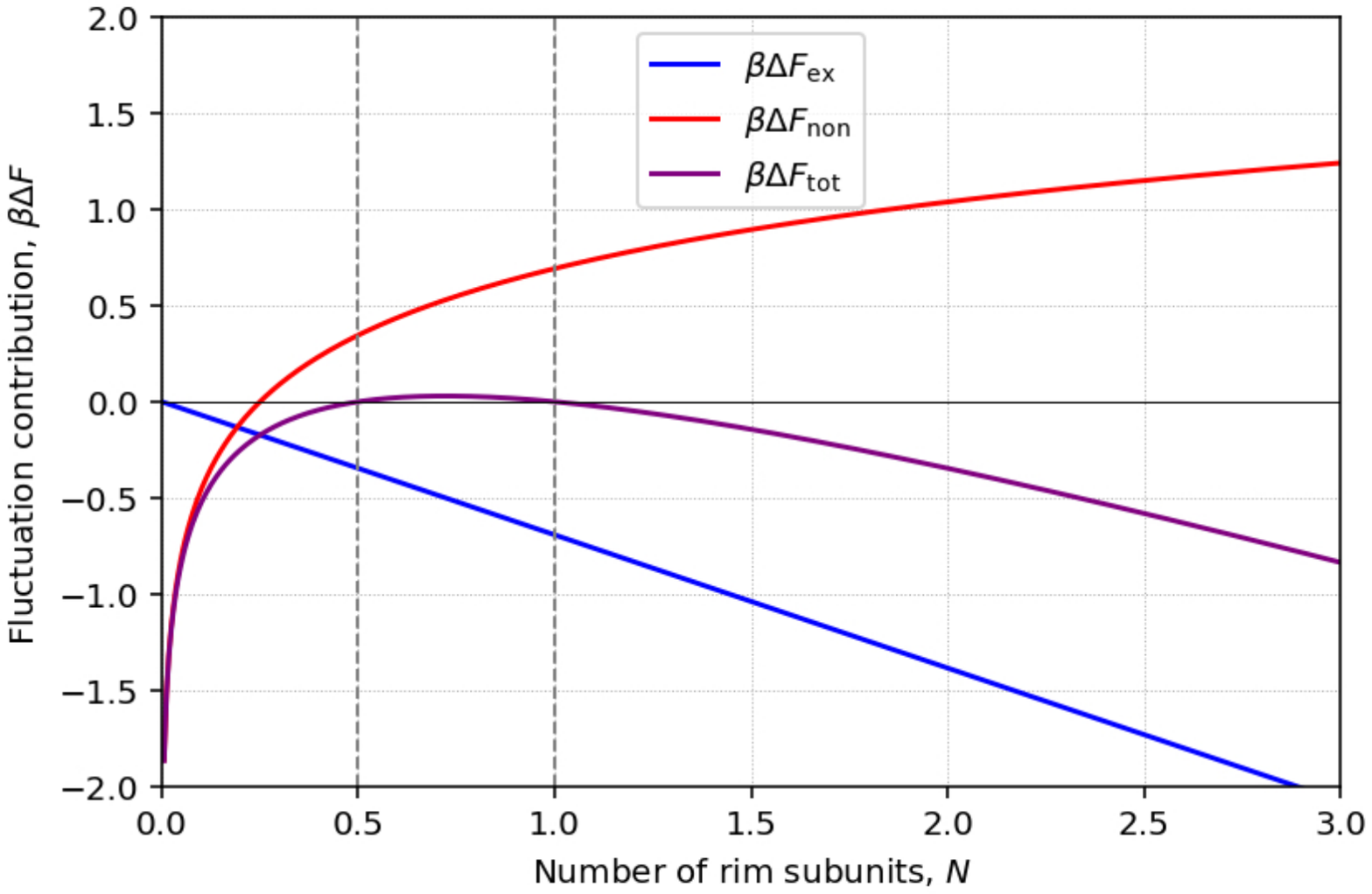}
    \caption{Dimensionless fluctuation free-energy contributions in the continuum rim model from Eq.~\eqref{eq:cont_deltaF} for the parameter value $\beta \varepsilon^{\text{C}}/2\pi=0.25<1/e$. The solid blue curve shows the extensive contribution $\beta \Delta F_{\mathrm{ext}}$, the solid red curve shows the non-extensive contribution $\beta \Delta F_{\mathrm{non}}$, and the solid purple curve shows the total fluctuation correction $\beta \Delta F$. The two dashed vertical gray lines mark the two real solutions of the zero-condition equation, corresponding to the two Lambert branches. In this regime, the fluctuation correction is positive only over a narrow interval of small rim sizes with $N_0<N<N_{-1}\leq 1$; since $N<1$ is non-physical, rim fluctuations lower the free energy for all physical rim sizes $N\geq1$.}
    \label{fig:case_3}
\end{figure}

\begin{figure}
    \centering
    \includegraphics[width=0.99\linewidth]{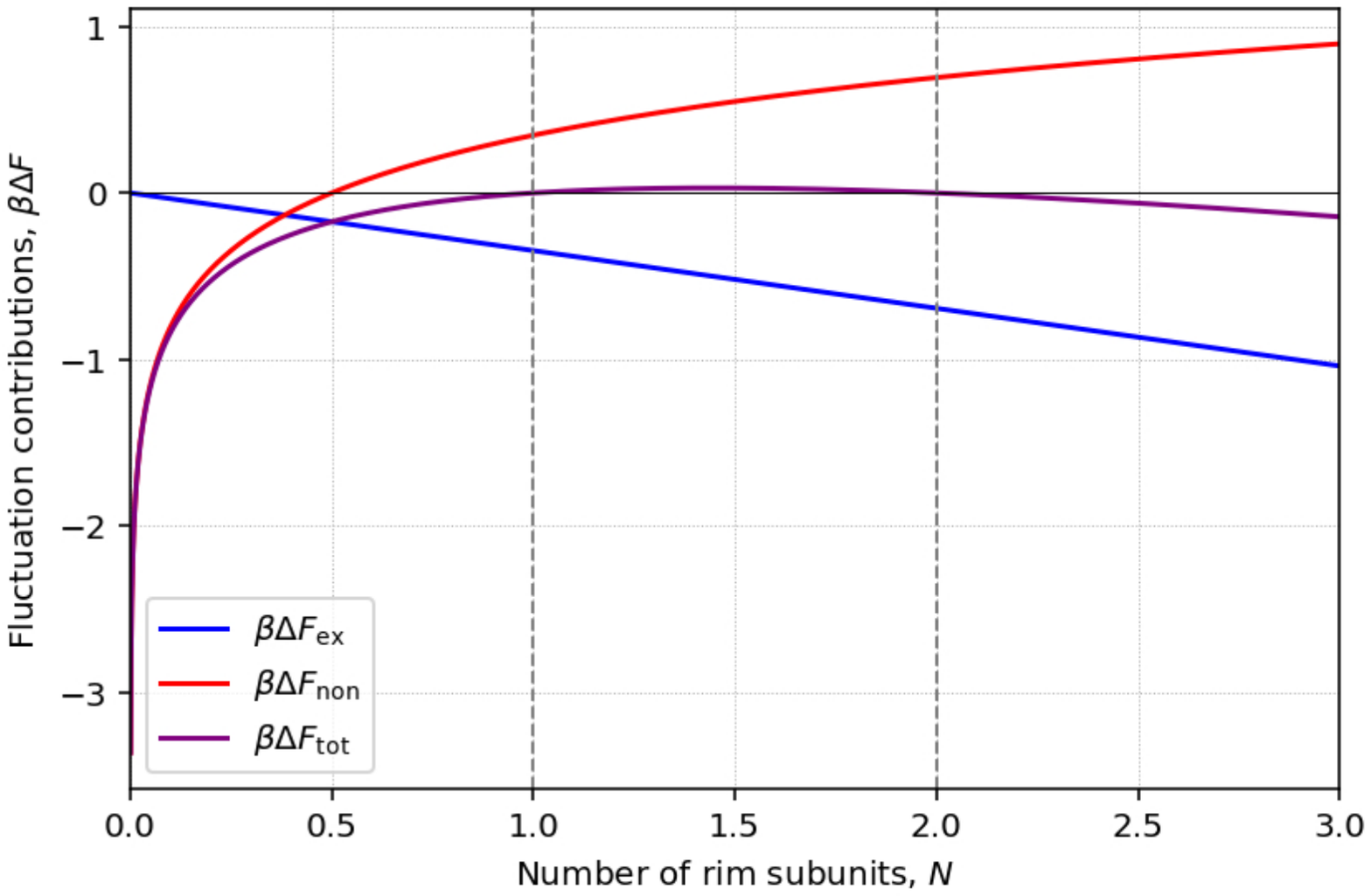}
    \caption{Fluctuation free-energy contributions in the continuum rim model from Eq.~\eqref{eq:cont_deltaF} as a function of the rim size $N$ for the intermediate parameter value $\beta \varepsilon^{\text{C}}/2\pi=0.5$, satisfying $1/e<\beta \varepsilon^{\text{C}}/2\pi<1$. The solid blue curve shows the extensive contribution $\beta \Delta F_{\mathrm{ext}}$, the solid red curve shows the non-extensive contribution $\beta \Delta F_{\mathrm{non}}$, and the solid purple curve shows the total fluctuation correction $\beta \Delta F$. The dashed vertical gray lines indicate the two real solutions of the zero-condition equation, which bound the finite interval $1<N<N_{-1}$ where $\beta \Delta F>0$. For larger rim sizes, the fluctuation correction becomes negative, lowering the nucleation barrier. The regime $N<1$ is unphysical, as the rim must consist of at least one bond.}
    \label{fig:case_4}
\end{figure}

Because $\beta \Delta F$ is a concave function of $N$, these two solutions define the lower and upper bounds of a finite interval of rim sizes for which the fluctuation correction is positive. Outside this interval, fluctuations reduce the free energy, as in the discrete model. Real solutions occur only when the argument of Eq.~\eqref{eq:lambert_definition} lies within the domain where both branches are defined. This behavior is illustrated in Figs.~\ref{fig:case_1}--\ref{fig:case_4}, where $\Delta F$ is positive only within the finite interval bounded by the two solutions $N_0$ and $N_{-1}$, and its magnitude in this regime remains modest. Note that values $N<1$ are non-physical, as the rim must consist of at least one bond; the analysis below is therefore restricted to the physical domain $N\geq 1$.

To illustrate, Fig.~\ref{fig:case_1} corresponds to the threshold case $\beta \varepsilon^{\text{C}}/2\pi=1/e$. In this limit, the two Lambert branches coalesce, and $\beta \Delta F$ touches zero only at $N=1$. Consequently, no finite window of positive $\beta \Delta F$ exists beyond this point. Since $\ln N = 0$ at $N=1$, this value always constitutes a boundary where $\beta \Delta F = 0$. 

Similarly, for $\beta \varepsilon^{\text{C}}/2\pi > 1/e$, illustrated in Fig.~\ref{fig:case_2}, there is only one real solution, corresponding to the principal branch $N_0=1$. For all physical rim sizes $N\geq 1$, the fluctuation contribution is positive, indicating that rim fluctuations uniformly raise the free energy barrier in this regime.

For the case $0 < \beta \varepsilon^{\text{C}}/2\pi < 1/e$, illustrated in Fig.~\ref{fig:case_3}, two real solutions exist. The principle branch yields $N_0<1$, while the lower branch gives $N_{-1}=1$. Because the positive $\Delta F$ window lies entirely within the interval $N_0<N<N_{-1}\leq 1$, it falls in the non-physical domain. Consequently, for all physical rim sizes $N\geq 1$, the fluctuation correction is slightly negative, and rim fluctuations uniformly lower the free energy.

Finally, for $1/e < \beta \varepsilon^{\text{C}}/2\pi < 1$, illustrated in Fig.~\ref{fig:case_4}, two distinct real solutions again exist, with $N_0=1$ and $N_{-1}>1$. Thus the fluctuation correction is positive for all physical rim sizes in the interval $1\leq N<N_{-1}$ and becomes negative for $N>N_{-1}$.

In physical terms, these results imply that when the nucleation barrier is low, either due to strong subunit binding or a small critical nucleus size, rim fluctuations can raise the barrier slightly for a finite range of partially assembled capsid sizes. Conversely, when the barrier is high, fluctuations uniformly lower the barrier. The continuum analysis thus confirms and extends the conclusions of the discrete model, providing a unified picture of how rim flexibility influences capsid assembly thermodynamics.

\section{Entropic Effects on Capsid Nucleation\label{Sec:Entropy}}

While the free--energy analysis already captures how rim fluctuations modify the nucleation landscape, examining the entropy explicitly allows us to separate configurational and energetic contributions and to assess the thermodynamic consistency of the fluctuation models. In this section, we compute the entropy associated with rim fluctuations and quantify their entropic contribution to capsid assembly.

The entropy is obtained from the Helmholtz free energy via
$S = -(\partial F/\partial T)_N$.
We ignore the temperature dependence of the binding free energy~\cite{Kegel_and_Schoot}, which enters the total free energy through the line tension $\gamma$~\cite{zandi_classical_2006}. As a result, the baseline rim energy $F_0$ is temperature independent and does not contribute to the entropy; all entropic effects therefore arise solely from the fluctuation degrees of freedom of the rim.

\subsection{Entropy of the three-state model}

Differentiating Eq.~(\ref{discrete F}) yields the entropy of the three-state rim model,
\begin{align}
    S=Nk_{\textsc{b}}\ln\!&\left[1+2e^{-\beta \varepsilon^{\text{D}}}\right] + \frac{k_{\textsc{b}}}{2}\ln\!\left[\frac{1+\tfrac12 e^{\beta \varepsilon^{\text{D}}}}{2\pi N}\right]\nonumber\\
    &+\frac{\gamma \Delta \ell}{T}\left(N-\frac{2N+1}{2(1+2e^{-\beta \varepsilon^{\text{D}}})}\right)
    \label{discrete entropy}
\end{align}
where $-\beta \varepsilon^{\text{D}}<0$.

The entropy separates naturally into three contributions. The first term is extensive in $N$ and reflects the configurational entropy associated with the multiplicity of allowed rim-step configurations. The second, logarithmic term is a subdominant finite-size correction arising from the exact rim-closure constraint. The final term arises from the temperature dependence of the Boltzmann weights of higher-energy rim configurations relative to the flat rim state.

In the strict low-temperature limit, the exact discrete model given in Eq.~\eqref{discrete Z} gives $S\to 0$; any nonzero residual obtained by extrapolating Eq.~\eqref{discrete entropy} reflects the breakdown of the saddle-point approximation outside of its validity regime (see the text following Eq.~\eqref{factored Z} and Appendix~\ref{appendix D}).

\subsection{Entropy of the continuous model}

In the continuum capillary-wave limit, differentiating Eq.~(\ref{continuum approx result}) yields
\begin{align}
S = k_{\textsc{b}}\frac{N-1}{2}
\left(
1+\ln\!\left[\frac{2\pi}{\beta\varepsilon^{\text{C}}}\right]
\right)
-\frac{k_{\textsc{b}}}{2}\ln N.
\label{continuum entropy}
\end{align}

The entropy consists of an extensive contribution proportional to $(N-1)/2$, reflecting the number of independent fluctuation modes of the rim, and a non-extensive logarithmic correction imposed by the global closure constraint.

Unlike the discrete model, however, the continuum expression does not remain non-negative for all parameters. For $N>1$, setting $S=0$ gives
\begin{align}
\frac{2\pi}{\beta \varepsilon^{\text{C}}}
=
\frac{N^{1/(N-1)}}{e},
\end{align}
so that $S<0$ when $\beta\varepsilon^{\text{C}}/2\pi>e/N^{1/(N-1)}$. In particular, $S\to -\infty$ as $T\to 0$, whereas for $N=1$ the entropy vanishes identically. Since $N<1$ is non-physical, the entropy expression is meaningful only for $N\geq 1$. The origin and interpretation of this behavior are discussed in the following section.

\section{Analysis and Discussion\label{Sec: Analysis and Discussion}}
\subsection{Nucleation barriers}
Based on classical nucleation theory, a partially formed capsid with $n$ subunits is described as a spherical cap bounded by a rigid, circular rim. Within this framework, the Gibbs free energy of formation is
\begin{align}
    \Delta G(n)\simeq -n\Delta\mu + \gamma L_n,
    \label{Delta G}
\end{align}
where $n$ denotes the number of coat protein subunits that make up the incomplete shell, $\Delta \mu$ is the chemical potential difference between a subunit in solution and a subunit in the fully assembled capsid, and $\gamma$ is the effective line tension associated with the open rim of the partial shell. The rim length of a spherical cap of cluster of size $n$ is
\begin{align}
    L_n=\frac{4\pi R}{q}\sqrt{n(q-n)},
\end{align}
with $R$ the capsid radius and $q$ the total number of subunits in the complete shell~\cite{zandi_classical_2006}.

The number of rim bonds is equal to $N=N(n)=L_n/a$, where $a$ is their microscopic spacing along the rim. For an icosahedral capsid, the total number of subunits is $q=60T$, where $T$ is the triangulation number~\cite{1962CK}. This gives
\begin{align}
    N(n)=\frac{L_n}{a}=\frac{4\pi R}{qa}\sqrt{n(q-n)}.
\end{align}

\subsubsection{Barriers in the Discrete Rim Model}
Using the rim free energy, Eq.~(\ref{discrete F}), we now calculate the nucleation barriers for capsid formation. The free energy of an $n$-subunit cluster in the discrete (three-state) rim model is
\begin{align}
\Delta G^{D}(n)
= -n\Delta\mu + F_{\mathrm{rim}}^{D}\!\left(N(n)\right),
\end{align}
where $N(n)$ is the number of bonds along the rim.  Introducing the effective line tension
\begin{align}\label{gDeff}
\gamma_{\mathrm{eff}}^{D}
= \gamma - \frac{1}{\beta a}
\ln\!\left[1+2e^{-\beta \varepsilon^{\mathrm{D}}}\right],
\end{align}
the rim free energy in Eq.~(\ref{discrete F}) can be written as
\begin{align}\label{FDrim}
F_{\mathrm{rim}}^{D}(N)
= \gamma_{\mathrm{eff}}^{D} Na
-\frac{1}{2\beta}
\ln\!\left[
\frac{1}{2\pi N}
\left(1+\frac{1}{2}e^{\beta \varepsilon^{\mathrm{D}}}\right)
\right],
\end{align}
where the first term in Eq.~\ref{FDrim} is extensive in $N$ and represents the dominant entropic renormalization of the line tension. The second term originates from the exact closure constraint on the rim and scales logarithmically with $N$; we denote this subdominant contribution by $\Delta F_{\mathrm{non}}(N)$.

The nucleation free energy may therefore be written as
\begin{align}
\Delta G^{D}(n)
\simeq -n\Delta\mu
+ \gamma_{\mathrm{eff}}^{D} L_n
+ \Delta F_{\mathrm{non}}\!\left(N(n)\right),
\end{align}
with the same geometric dependence of $L_n$ on $(n,q)$ as in the classical theory.

For supersaturations $\Delta\mu$ such that a finite nucleation barrier exists within the capillarity approximation, the reduced effective line tension $\gamma_{\mathrm{eff}}^{D}<\gamma$ lowers the barrier and shifts the critical nucleus size $n^\ast$ associated with the maximum of the free energy $\Delta G$ to smaller values compared to the rigid-rim case. As discussed earlier, the finite-size correction associated with rim closure can be positive over a limited range of $N$, producing a narrow regime in which fluctuations slightly increase the free energy of very small nuclei.

\subsubsection{Barriers in the Continuum Limit}
Using the continuum fluctuation free energy, Eq.~(\ref{eq:cont_deltaF}), we now calculate the nucleation barriers. Adding the fluctuation correction to the classical rim energy $F_0$, the rim free energy can be written as
\begin{align}
F_{\mathrm{rim}}^{C}(N)
= {F_0} + \Delta F_{\mathrm{cont}}(N),
\end{align}
and the free energy of an $n$-subunit nucleus becomes
\begin{align}
\Delta G^{C}(n)
= -n\Delta\mu + F_{\mathrm{rim}}^{C}\!\left(N(n)\right).
\end{align}

As in the discrete model, the extensive part of the fluctuation correction renormalizes the effective line tension. Defining
\begin{align}\label{gCeff}
\gamma_{\mathrm{eff}}^{C}
= \gamma - \frac{1}{2\beta a}
\ln\!\left[\frac{2\pi}{\beta\varepsilon^{\text{C}}}\right],
\end{align}
the rim free energy may be expressed as
\begin{align}
F_{\mathrm{rim}}^{C}(N)
= \gamma_{\mathrm{eff}}^{C} Na
+ \Delta F_{\mathrm{non}}(N),
\end{align}
where the non-extensive correction arising from the closure constraint is given explicitly by
\begin{align}\label{nonextensiveF}
\Delta F_{\mathrm{non}}(N)
= -\frac{1}{2\beta}
\ln\!\left[\frac{\beta\varepsilon^{\text{C}}}{2\pi N}\right].
\end{align}

The nucleation free energy can therefore be written as
\begin{align}
\Delta G^{C}(n)
\simeq -n\Delta\mu
+ \gamma_{\mathrm{eff}}^{C} L_n
+ \Delta F_{\mathrm{non}}\!\left(N(n)\right).
\end{align}

For parameter values $\beta\varepsilon^{\text{C}}$ of order unity, continuum rim fluctuations substantially reduce the effective line tension, leading to a pronounced decrease in the nucleation barrier compared to the rigid-rim model. The logarithmic correction provides a finite-size contribution that is most relevant for small clusters.

\subsubsection{Effects on Barrier Height, Critical Nucleus, and Line Tension}
Equations~\eqref{gDeff}, \eqref{gCeff}, and \eqref{generalized effective gamma} show that rim fluctuations modify the classical nucleation free energy through a rim-dependent correction to the line-tension term. In all cases, the dominant contribution of this correction is extensive in $N$ and can be interpreted as a renormalization of the line tension. The nucleation free energy may therefore be described in terms of an effective line tension,
\begin{align}
\gamma_{\mathrm{eff}}
= \gamma + \frac{1}{L_n}\,\Delta F_{\mathrm{fluc}}\!\left(N(n)\right).
\end{align}

If $\Delta F_{\mathrm{fluc}}$ is negative, $\gamma_{\mathrm{eff}}$ decreases, which lowers the nucleation barrier height $\Delta G(n^\ast)$ and shifts the critical nucleus size $n^\ast$ to smaller values relative to the rigid-rim prediction. Within the classical nucleation theory, the critical nucleus size is given by
\begin{align}
    n^\ast = \frac{q}{2}\left(1-\frac{\alpha}{\sqrt{4+\alpha^2}}\right),
\end{align}
with $\alpha = q\Delta\mu/2\pi R\gamma_{\mathrm{eff}}$. Subdominant logarithmic corrections associated with rim closure, given in Eq.~(\ref{nonextensiveF}), lead to modest additional shifts in $n^\ast$ and can produce narrow intervals in $N$ for which the net fluctuation contribution is positive. In these regimes, rim fluctuations may slightly increase the free energy of very small clusters, delaying the onset of downhill growth to larger $n$.

From the perspective of classical nucleation theory, the line tension $\gamma$ is assumed to be a constant, independent of cluster size or rim geometry. The fluctuating rim models considered here retain the same microscopic line tension $\gamma$, but show that the thermodynamic contribution of the interface to the free energy is renormalized by entropy. As a result, the effective line tension $\gamma_{\mathrm{eff}}$ depends on temperature, microscopic geometry, and the spectrum of allowed rim fluctuations, and may acquire a weak dependence on cluster size through finite-size corrections. In this sense, rim fluctuations refine the capillarity approximation by replacing a strictly constant line tension with an effective description that incorporates both extensive and finite-size entropic contributions.

\section{Conclusion}

We have shown that incorporating rim fluctuations into the nucleation theory of spherical capsids can modify both the nucleation barrier and the critical nucleus size. The theory makes no assumption about capsid symmetry and is therefore applicable to both icosahedral and non-icosahedral spherical viruses \cite{li2022biophysical}.  In classical nucleation theory, the barrier arises from a competition between bulk binding free energy and a fixed line tension associated with a rigid circular rim. Allowing the rim to undergo geometric fluctuations introduces an entropic contribution that renormalizes the effective line tension, along with a subdominant finite-size correction associated with rim closure.

In most parameter regimes relevant for assembly, the extensive entropic contribution dominates and lowers the nucleation barrier, thereby facilitating capsid formation. Under certain intermediate conditions, however, the closure constraint can partially offset this entropy gain, leading to a modest increase in the free energy of very small clusters. Importantly, rim fluctuations do not alter the overall chemical potential difference $\Delta\mu$, which continues to control the equilibrium between assembled and disassembled states. Instead, fluctuations primarily couple to rim geometry and to the energetic cost of deviations from the minimal rim length, reshaping the free-energy landscape of intermediate assembly states.

A direct comparison of the discrete and continuum descriptions shows that both frameworks capture the same underlying entropic mechanism. In each case, rim fluctuations generate a dominant extensive entropy that lowers the effective line tension, together with a smaller logarithmic correction associated with rim closure. The discrete formulation remains thermodynamically well behaved as $T \to 0$, satisfying the Third Law. In the continuum model, the negative entropy obtained in the low-temperature limit should be understood as an artifact of continuum coarse-graining. The free-energy functional is defined relative to a reference state and integrates out microscopic degrees of freedom, leading to an effective entropy that is not constrained to be positive. This does not constitute a violation of the third law of thermodynamics, which applies to the absolute entropy of the complete microscopic system. In practice, the continuum description is not expected to remain valid arbitrarily close to $T=0$, where discreteness and microscopic effects become essential.

In the discrete rim model, the energetic penalty for rim deformations is parametrized by the geometric difference $\Delta \ell$, which quantifies the deviation of a rim segment from its minimal configuration and sets the amplitude of allowed fluctuations. In this sense, the energetic and entropic contributions to the rim free energy are not independent but are coupled through the geometry of the fluctuating boundary. Larger-amplitude rim fluctuations simultaneously increase configurational entropy and incur a higher energetic cost, and the balance between these effects determines the effective line tension and the stability of intermediate assembly states.

These results demonstrate that the entropic reduction of the effective rim tension is a robust feature that does not depend on the microscopic details of the rim model. Instead, it reflects a general consequence of allowing boundary fluctuations in partially formed shells. As a result, rim fluctuations lead to a threshold behavior in the subunit--subunit binding energy: below a threshold value, fluctuations reduce the nucleation barrier and promote assembly, whereas for sufficiently strong binding energies they can stabilize incomplete capsids by increasing the barrier.

More broadly, these findings suggest that rim fluctuations play an increasingly important role for larger capsids, which support a greater number of fluctuation modes. They may also help rationalize the frequent involvement of scaffolding proteins or packaged genomes in large viral assemblies, as such additional constraints are expected to suppress rim fluctuations, reduce the associated configurational entropy, increase the effective line tension $\gamma_{\mathrm{eff}}$, and raise the nucleation barrier. By explicitly incorporating boundary fluctuations, this work extends classical nucleation theory and provides a more complete physical framework for understanding nucleation and assembly in protein shells.

\acknowledgements{We dedicate this work to the memory of Rudolf Podgornik, a long-time collaborator and friend. We benefited greatly from his deep physical insight, intellectual generosity, and many stimulating discussions over the years. His influence on this work, and on the field more broadly, will continue to be felt.  This work was supported by the National Science Foundation under Grant Nos. NSF DMR-2131963 and
MCB/PHY-2413062.}

\section*{Author Declarations}
\subsection*{Conflict of Interest}
The authors have no conflicts to disclose.

\subsection*{Author Contributions}
\noindent \textbf{Alexander Bryan Clark:} Data curation (equal); Formal analysis (equal); Investigation (lead); Methodology (equal); Software (lead); Validation (equal); Visualization (lead); Writing - original draft (lead); Writing - review \& editing (equal).
\textbf{Paul van der Schoot:} Conceptualization (equal); Data curation (equal); Formal analysis (equal); Investigation (equal); Methodology (equal); Validation (equal); Visualization (equal); Writing - original draft (equal); Writing - review \& editing (equal).
\textbf{Henri Orland:} Conceptualization (equal); Data curation (equal); Formal analysis (equal); Investigation (equal); Methodology (equal); Validation (equal); Visualization (equal); Writing - original draft (equal); Writing - review \& editing (equal).
\textbf{Roya Zandi:} Conceptualization (equal); Data curation (equal); Formal analysis (equal); Funding acquisition (lead); Investigation (equal); Methodology (equal); Project administration (lead); Supervision (lead); Validation (equal); Visualization (equal); Writing - original draft (equal); Writing - review \& editing (equal).

\section*{Data Availability}
The data that supports the findings of this study are available upon request. 

\appendix

\section{Generalized Discretization of Fluctuations on an Arc-Length\label{Appendix A}}
\begin{figure}
    \centering
    \includegraphics[width=0.99\linewidth]{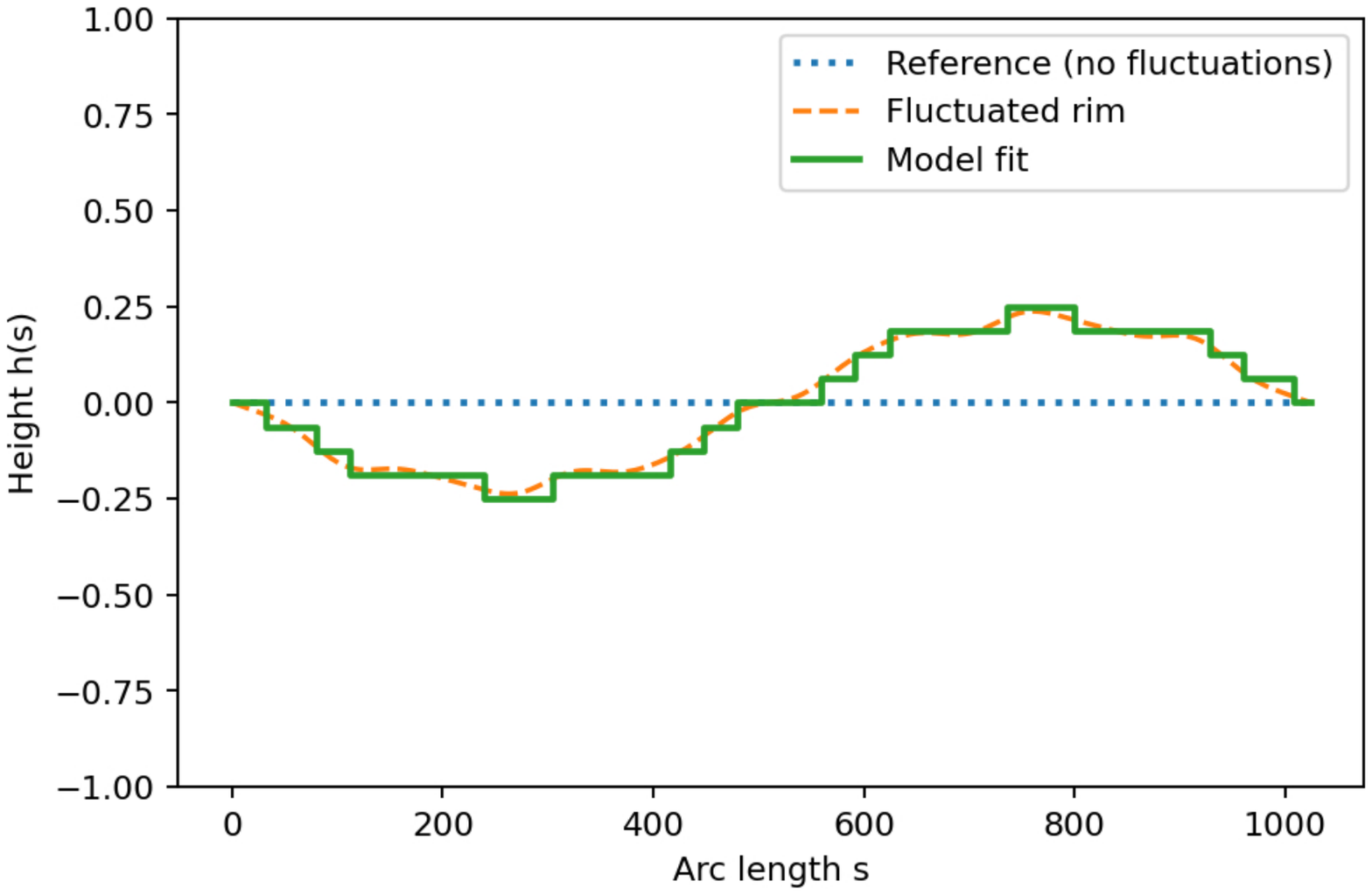}%
    \caption{A generalized multi-state discrete step model for rim fluctuations. In this model, a rim bond can have a step height that is an integer multiple of $|b|$. A randomly generated rim height profile $h(s/L)$, represented by an orange dashed line, is shown along with the best fit of a $(2m{+}1)$-state discretized fluctuation model (solid green line). The blue dotted line is the reference rim with no fluctuations. \label{generalized_arc_length_fit}}%
\end{figure}

We generalize the three-state step model of the rim to allow multiple discrete step heights per bond. Consider a rim consisting of $N$ bonds of horizontal length $a$. Each bond $j$ may take an integer step 
\begin{align*}
    r_j\in\{-m,-m+1,\cdots,m-1, m\},
\end{align*}
for $m\in \mathbb{N}$, corresponding to a vertical displacement $r_jb$, where $b$ is the fundamental step height. See Fig.~\ref{generalized_arc_length_fit} for illustration.

The exact energy of a single bond with step $r$ is 
\begin{align}
    E_r=\gamma\sqrt{a^2+(rb)^2}
\end{align}
where $\gamma$ is the microscopic line tension.

For small slopes $b\ll a$, expanding the square root gives 
\begin{align}
    E_r = \gamma a +\frac{\gamma a r^2\lambda^2}{2},\label{Erj}
\end{align}
where we have kept up to the quadratic order in $r$.

\subsection{Partition Function and Free Energy}
A closed rim requires the net vertical displacement to vanish, $\sum_{j=1}^Nr_j=0$. The configurational partition function is therefore
\begin{align}
    Z_m=\sum_{\{r_j\}}\exp\left[-\beta \sum_{j=1}^N E_{r_j}\right]\delta\left(\sum_{j=1}^Nr_j\right).\label{Zm with delta}
\end{align}
Using the Fourier representation of the Kronecker delta on integers and substituting the quadratic approximation Eq.~\eqref{Erj}, we obtain
\begin{align}
    Z_m = e^{-\beta \gamma Na}\int_{-\pi}^\pi \frac{\mathrm{d}k}{2\pi} \left(Z_1^{(m)}(k)\right)^N, \label{Zm with 1bond factor}
\end{align}
with the single-bond partition function
\begin{align}
    Z_1^{(m)}(k)=\sum_{r=-m}^m \exp\left[-\frac{1}{2} \beta\gamma a r^2\lambda^2 +ikr \right]. \label{1bond factor}
\end{align}

For large $N$, the integral in Eq.~\eqref{Zm with 1bond factor} can be evaluated by steepest descent. By symmetry, $k=0$ is an extremum of $Z_1^{(m)}(k)$, and it is the dominant contribution in the present model. Defining the $k=0$ single-bond partition as $Z_1^{(m)}$, we have
\begin{align}
    Z_1^{(m)}=\sum_{r=-m}^m \exp\left[-\frac{1}{2}\beta \gamma a r^2\lambda^2\right].\label{single bond at k=0}
\end{align}
This now allows us to introduce averages with respect to the normalized single-bond weights,
\begin{align}
    \left<f(r)\right>=\frac{1}{Z_1^{(m)}}\sum_{r=-m}^m f(r) \exp\left[-\frac{1}{2}\beta \gamma a r^2\lambda^2\right].
\end{align}
Expanding $\ln\left[Z_1^{(m)}(k)\right]$ about $k=0$ gives
\begin{align}
    \ln\left[Z_1^{(m)}(k)\right] = \ln\left[Z_1^{(m)}\right] -\frac{1}{2}Ak^2,\label{Zm1 expansion}
\end{align}
with curvature, $A$, given by
\begin{align}
    A \equiv \left.-\frac{\mathrm{d}^2\ln\left[Z_1^{(m)}\right]}{\mathrm{d}k^2}\right|_{k=0}
\end{align}
which is exactly the variance of $r$. Therefore, $A=\left<r^2\right>-\left<r\right>^2$. Since the allowed steps are symmetric, $\left<r\right>=0$, hence $A=\left<r^2\right>$.

Substituting Eq.~\eqref{Zm1 expansion} into Eq.~\eqref{Zm with 1bond factor} and extending the Gaussian approximation yields
\begin{align}
    \int_{-\pi}^\pi\frac{\mathrm{d}k}{2\pi}\exp\left[N\ln\left[Z_1^{(m)}(k)\right]\right]\approx \frac{\left(Z_1^{(m)}\right)^N}{\sqrt{2\pi N\left<r^2\right>}}.
\end{align}
Therefore, the large $N$ partition function is 
\begin{align}
    Z_m \approx e^{-\beta \gamma Na}\frac{\left(Z_1^{(m)}\right)^N}{\sqrt{2\pi N\left<r^2\right>}}.
\end{align}
Implying that the dimensionless free energy is 
\begin{align}
    \beta F_m = \beta \gamma Na - N\ln\left[Z_1^{(m)}\right]+\frac{1}{2}\ln\left[2\pi N\left<r^2\right>\right]. \label{generalized free energy}
\end{align}
It is convenient to separate the extensive part from the closure-induced non-extensive correction by defining an effective line tension
\begin{align}
    \gamma_{\mathrm{eff}}^{(m)} \equiv \gamma -\frac{1}{\beta a}\ln\left[Z_1^{(m)}\right],\label{generalized effective gamma}
\end{align}
so that 
\begin{align}
    \beta F_m = \beta \gamma_{\mathrm{eff}}^{(m)}Na +\frac{1}{2}\ln\left[2\pi N\left<r^2\right>\right].\label{generalized effective F}
\end{align}
The last term is the finite-$N$ contribution generated by the exact closure constraint.

It is also worth noting that in the thermodynamic limit $N\to\infty $
\begin{align}
    \lim_{N\to\infty}\frac{\beta F_m}{N}=\beta \gamma_{\mathrm{eff}}^{(m)} a
\end{align}
and the closure term vanishes per bond.

\subsection{Continuum Limit of the Generalized Discrete Model}
In the regime where the step spectrum becomes dense, an approximation can be made to the discrete sum in Eq.~\eqref{single bond at k=0} by a Gaussian integral. Formally, for large $m$ and small step size such that the relevant contributing steps satisfy $|r|\ll m$, the following replacement can be made
\begin{align}
    Z_1^{(m)} &\approx \int_{-\infty}^\infty \mathrm{d}r\, \exp\left[-\frac{1}{2}\beta \gamma ar^2\lambda^2\right],\\
    & = \sqrt{\frac{2\pi}{\beta \varepsilon^{\text{C}}}}. \label{generalized approx 1}
\end{align}
Similarly, 
\begin{align}
    \left<r^2\right> &\approx \frac{\int \mathrm{d}r\, r^2\exp\left[-\frac{1}{2}\beta\gamma ar^2\lambda^2 \right]}{\int \mathrm{d}r\, \exp\left[-\frac{1}{2}\beta\gamma ar^2\lambda^2 \right]},\\
    &= \frac{1}{\beta \varepsilon^{\text{C}}} \label{generalized approx 2}.
\end{align}
Substituting Eq.~\eqref{generalized approx 1} and Eq.~\eqref{generalized approx 2} into Eq.~\eqref{generalized free energy} reproduces the continuum capillary-wave free energy of the main text (up to the usual considerations regarding microscopic cutoffs).

\subsection{Entropy of the Generalized Discrete Model}
The entropy follows from $S_m=-(\partial F_m/\partial T)_N$. Using Eq.~\eqref{generalized free energy}, and the $\beta$-dependence of $Z_1^{(m)}$ and $\left<r^2\right>$, one finds
\begin{align}
    S_m = k_{\textsc{b}}N&\ln\left[Z_1^{(m)}\right] - \frac{k_{\textsc{b}}}{2}\ln\left[2\pi N\left<r^2\right>\right]\nonumber\\
    &+ \frac{\varepsilon^{\text{C}}}{2T}\left(N\left<r^2\right>-\frac{1}{2}\frac{\left<r^4\right>-\left<r^2\right>^2}{\left<r^2\right>}\right).
\end{align}
For any fixed finite $m$, the model is thermodynamically well-behaved as $T\to 0$ as the distribution collapses to $r=0$ and $S_m\to0$. In the dense-spectrum or continuum limit, the familiar low-temperature pathology of the naive continuum approximation reappears unless a microscopic cutoff is retained (see Appendix~\ref{appendix D}).

\subsection{Nucleation Barriers in the Generalized Discrete Model}
Let $N(n)$ denote the number of rim bonds for a nucleus of size $n$ (as defined in the main text). The nucleation free energy in the generalized $(2m+1)$-state rim model is 
\begin{align}
    \Delta G^{(m)}(n) =- n\Delta \mu + F_m(N(n)),
\end{align}
with $F_m$ given in Eq.~\eqref{generalized free energy}, or equivalently, Eq.~\eqref{generalized effective F}. In particular,
\begin{align}
    \Delta G^{(m)}(n)=-n\Delta\mu +\gamma_{\text{eff}}^{(m)}N(n)a+\frac{1}{2\beta}\ln\left[2\pi N(n)\left<r^2\right>\right],
\end{align}
implying that an increase in $m$ increases the number of available step states, raises the configurational entropy, and typically lowers $\gamma_{\mathrm{eff}}^{(m)}$.

\section{Technical Steps in the Continuum Limit\label{appendix continuous}}
Starting from the quadratic partition function of the continuum rim model, Eq.~\eqref{continuum Z}, the constrained Gaussian weight can be written as
\begin{align}
    \exp\left[-\beta \gamma a\lambda^2\sum_{k,l=2}^N n_kM_{kl}n_l\right],
\end{align}
where $M$ is the $(N-1)\times(N-1)$ tridiagonal matrix
\begin{align}
    M=\left(\begin{matrix}
         1 & -\frac{1}{2} & \cdots & 0 & 0  \\
         -\frac{1}{2} & 1 & -\frac{1}{2} & \cdots & 0\\
         \cdots & -\frac{1}{2} & 1 & -\frac{1}{2} & \cdots \\
         0 & \cdots & -\frac{1}{2} & 1 & -\frac{1}{2}\\
         0 & 0 & \cdots & -\frac{1}{2} & 1
    \end{matrix}\right).
\end{align}

The Gaussian integral yields
\begin{align}
    Z = e^{-\beta \gamma Na }\left(\frac{\pi}{\beta \gamma a\lambda^2}\right)^{(N-1)/2} \left(\det{M}\right)^{-1/2}.\label{Z with determinant}
\end{align}
The determinant obeys the recursion relation yielding $\det M = N/2^{N-1}$. Substituting the determinant into Eq.~\eqref{Z with determinant} gives the dimensionless free energy 
\begin{align}
    \beta F=\beta \gamma Na-\frac{N-1}{2}\ln\left[\frac{2\pi}{\beta \gamma a\lambda^2}\right] +\frac{1}{2}\ln N.
\end{align}


\section{Free Energy of Fluctuations\label{appendix C}}

Consider a system initially constrained to a single state with energy $E_0$. Adding higher-energy configurations $\{E_i\}$ with $E_i \geq E_0$ changes the partition function from $Z_0 = e^{-\beta E_0}$ to
\begin{align}
    Z = e^{-\beta E_0}\left(1 + \sum_i e^{-\beta (E_i - E_0)}\right).
\end{align}
Since the term in parentheses exceeds unity, the free energy
$F = -(1/\beta)\ln Z < E_0$, demonstrating that increasing the number of accessible configurations always lowers the free energy. The corresponding reduction is
\begin{align}
    \Delta F = -\frac{1}{\beta}\ln\left[1 + \sum_i e^{-\beta (E_i - E_0)}\right],
\end{align}
which implies $\Delta F < 0$. This principle underlies the entropic reduction of the effective line tension in all rim fluctuation models. Conversely, imposing constraints reduces the number of accessible states and raises the free energy, as seen in the closure constraint contribution $\Delta F_{\mathrm{non}} > 0$ discussed in the main text.

It should be emphasized that the above reasoning applies strictly to systems with a discrete set of states. For systems with continuous degrees of freedom, the sum over configurations must be replaced by an integral weighted by an appropriate density of states. In that case, Eqs.~(C1) and (C2) do not apply directly, and the resulting entropy depends on the chosen measure and coarse-graining procedure.

\section{The Difference Between Summing and Integrating\label{appendix D}}
Replacing a discrete sum over states by a continuum integral can change the thermodynamic measure and yield nonphysical low-temperature behavior.

Consider energies $E_n=E_0+\alpha n^2$ for integer $n = 0, \pm1, \pm2, \dots$, and $\alpha>0$. The exact discrete partition function is 
\begin{align}
    Z_{\text{sum}}&= e^{-\beta E_0}\sum_{n=-\infty}^\infty \exp[-\beta\alpha n^2]
\end{align}
which may be rewritten in terms of the Jacobi theta function
\begin{align}
     Z_{\text{sum}}&=e^{-\beta E_0}\vartheta_3\left(0,e^{-\beta \alpha}\right).
\end{align}
The discrete free energy is therefore
\begin{align}
    F_{\text{sum}}=E_0-\frac{1}{\beta}\ln\left[\vartheta_3\left(0,e^{-\beta \alpha}\right)\right].
\end{align}
For any $T>0$, the Jacobi theta function is greater than 1 for $0<e^{-\beta \alpha}<1$, hence $F_{\text{sum}}<E_0$. Moreover, as $T\to 0$, $\vartheta_3(0,e^{-\beta \alpha})\to 1$ so $F_{\text{sum}}\to E_0$ and $S\to 0$, consistent with the third law.

A naive continuum replacement treats $n$ as continuous and approximates 
\begin{align}
    Z_{\text{int}}&=e^{-\beta E_0}\int_{-\infty}^{\infty}\mathrm{d}n\ \exp\left[-\beta \alpha n^2\right],\\
    &= e^{-\beta E_0}\sqrt{\frac{\pi}{\beta \alpha}}.
\end{align}
The corresponding free energy is 
\begin{align}
    F_{\text{int}} = E_0-\frac{1}{2\beta}\ln\left[\frac{\pi}{\beta\alpha}\right].
\end{align}
For sufficiently large $\beta$, $\ln(\pi/\beta\alpha)<0$, implying $F_{\mathrm{int}}>E_0$. The entropy inferred from this continuum free-energy expression then becomes negative and diverges as $T \to 0$. This behavior reflects the coarse-grained nature of the continuum description: the free-energy functional is defined relative to a reference state and excludes microscopic degrees of freedom, resulting in an effective entropy that need not remain positive.

The implication for rim fluctuation theories is that entropies obtained from continuum mode integrals should be interpreted as effective, relative quantities. In particular, their low-temperature behavior signals the breakdown of the continuum approximation, rather than any violation of thermodynamic principles.

\bibliography{ref.bib}

\end{document}